\documentclass[showpacs,preprintnumbers,prd,amssymb,nofootinbib]{revtex4}
\RequirePackage[]{amsmath}
\RequirePackage[]{graphics}

\def\({\left(}
\def\){\right)}
\def\ls{\left[}
\def\rs{\right]}

\newcommand{\bea}{\begin{eqnarray}}
\newcommand{\eea}{\end{eqnarray}}
\newcommand{\nn}{\nonumber \\}
\newcommand{\bean}{\begin{eqnarray*}}
\newcommand{\eean}{\end{eqnarray*}}

\newcommand{\n}[1]{\label{#1}}

\newcommand{\Ref}[1]{(\ref{#1})}

\begin{document}

\title{Wormholes supported by the kink-like configuration of a scalar field}
\author{Sergey V. Sushkov}\email{sergey_sushkov@mail.ru}
\affiliation{Department of Mathematics, Kazan State Pedagogical
University, Mezhlauk 1 st., Kazan 420021, Russia}
\author{Sung-Won Kim}\email{sungwon@mm.ewha.ac.kr}
\affiliation{Department of Science Education, Ewha Womans
University, Seoul 120-750, Korea}


\begin{abstract}
We study the problem of existence of static spherically symmetric
wormholes supported by the kink-like configuration of a scalar
field. With this aim we consider a self-consistent, real,
nonlinear, nonminimally coupled scalar field $\phi$ in general
relativity with the symmetry-breaking potential $V(\phi)$
possessing two minima. We classify all possible field
configurations ruling out those of them for which wormhole
solutions are impossible. Field configurations admitting wormholes
are investigated numerically. Such the configurations represent a
spherical domain wall localized near the wormhole throat.
\end{abstract}

curved spacetime

\maketitle

\section{Introduction}
One of the most central feature of wormhole physics is the fact
that traversable wormholes as solutions to the Einstein equations
are accompanied by unavoidable violations of many of the energy
conditions, i.e., the matter threading the wormhole's throat has
to be possessed of ``exotic'' properties \cite{MT} (see also Ref.
\cite{HV}). The known classical matter does satisfy the usual
energy conditions, hence wormholes, if they exist, should arise as
solutions of general relativity and ``unusal" and/or quantum
matter. One approach is to regard wormholes as semiclassical in
nature \cite{Su92,HPS,PS,Pop01,THS,Od,Kha}. In particular,
self-consistent semiclassical wormhole solutions have been found
numerically in Refs. \cite{Su92,HPS}. Another way is to consider
``exotic'' field theories or theories based on various
modifications of general relativity. For example, wormholes
supported by ghost fields or fields with a negative sign kinetic
term was discussed in Refs.
\cite{Bron73,Kodama,HKL,Hayward,Armen}, and wormhole solutions in
the Brans-Dicke theory was studied in Refs.
\cite{Agnese,Nandi,Anch,HK}. It is worth to note that most of
these works can be considered with the unified point of view as
dealing with the scalar-tensor theory (STT) of gravity
\cite{Bron73,Bron1,BS,Bron-CC1,Bron-CC2}. The general STT action
is given by\footnote{Throughout this paper we use units such that
$G=c=1$.  The metric signature is $(- + + +)$ and the conventions
for curvature tensors are $R^\alpha_{\beta\gamma\delta} =
\Gamma^\alpha_{\beta\delta,\gamma} - ...$ and $ R_{\mu\nu} =
R^\alpha_{\mu\alpha\nu}$.}
\begin{equation}\n{action}
S=\int d^4x\sqrt{-g}\left\{\frac1{8\pi}f(\phi)R
-h(\phi)g^{\mu\nu}\phi_{,\mu}\phi_{,\nu}-2V(\phi)\right\},
\end{equation}
where $g_{\mu\nu}$ is a metric, $g=\det(g_{\mu\nu})$, $R$ is the
scalar curvature, $\phi$ is a scalar field, $f$ and $h$ are
certain functions of $\phi$, varying from theory to theory, and
$V(\phi)$ is a potential. The particular choice $h(\phi)\equiv-1$
in the action \Ref{action} leads to the class of theories with the
negative kinetic term. The Brans-Dicke theory \cite{BD}, described
in the Jordan frame, corresponds to the following case:
\begin{equation}f(\phi)=\phi,\quad h(\phi)=\frac{\omega}{\phi}\, ;\quad
\omega={\rm const}\not=\frac32.
\end{equation}

A ``less exotic'' theory is represented by the nonminimally
coupled scalar field in general relativity with the functions
\begin{equation}\n{xi}
f(\phi)=1-8\pi\xi\phi^2,\quad h(\phi)\equiv1;\quad \xi={\rm
const}.
\end{equation}
It turns out that such the theory also admits wormhole solutions.
For the case $V=0$ (massless scalar field) the static spherically
symmetric wormhole solutions was found in Ref. \cite{Bron73} (and
recently discussed in Ref. \cite{BV1}) for conformal coupling,
$\xi=1/6$, and in Ref. \cite{BV2} for any $\xi>0$. It is important
that more recent investigation \cite{BG} has shown that such the
solutions, obtained for the theory \Ref{action}, \Ref{xi} with
$V=0$, are unstable under spherically symmetric perturbations.

In this paper we study the problem of existence of wormhole
solutions in the theory \Ref{action}, \Ref{xi} with the
symmetry-breaking potential $V(\phi)$ possessing two minima. The
motivation for such the consideration is at least twofold. First,
there exists a common opinion that the spacetime of the early
Universe may have the nontrivial topological structure called the
spacetime foam. Such the ``foamy'' structure consists of objects
like primordial wormholes which can survive in the course of
cosmological evolution.\footnote{In the framework of the effective
action approach primordial wormholes, which induced from GUTs in
the early Universe, have been studied in Refs. \cite{Od}.} As is
well known the evolution of the early Universe is accompanied by
symmetry-breaking phase transitions which may have generated
nontrivial topologically stable field configurations (topological
defects) such as domain walls, strings, or monopoles
\cite{Kibble,VSh}. The interplay of gravity and nonlinear field
matter leads to a wealth of interesting phenomena \cite{remark}.
In particular, a static spherically symmetric wormhole solution
was found in the framework of the so-called $\lambda\phi^4$ model
with the ``abnormal'' negative-sign kinetic term \cite{Kodama}.
Also, the properties of spherical domain walls in a wormhole
spacetime have been studied in Ref. \cite{Su01}. The second
important reason, justifying our interest in this problem, is the
following: A scalar field theory with the symmetry-breaking
potential with two minima admits the existence of topologically
stable kink-like configurations (domain walls or interfaces), such
that the scalar field $\phi$ varies from $\phi_1$ to $\phi_2$,
where $\phi_{1,2}$ are vacuum states corresponding to the minima
of potential. Suppose that there exists a static spherically
symmetric wormhole solution supported by the kink-like
configuration of the scalar field, i.e., the spacetime geometry
represents two asymptotically flat regions (universies) connected
by a throat, and the scalar field $\phi$ is equal to $\phi_1$ in
one asymptotically flat region and $\phi_2$ in the other one,
varying substantially within a region near the throat. Such the
configuration, if it exists, should be topologically stable, and
so one may expect that it would also be stable under spherically
symmetric perturbations.

This article is organized as follows: In the section II we derive
the basic field equations. In the section III we, analyzing the
field equations, represent some general analytical results and
theorems and classify all possible field configurations ruling out
that of them for which wormhole solutions are impossible. The
particular model of the scalar field with the symmetry-breaking
potential is considered in the section IV. We analyze this model
numerically and obtain solutions describing wormholes supported by
the kink-like configuration of the scalar field. Some discussion
is given in the section V.

\section{Field equations}

Consider a real scalar field $\phi$ nonminimally coupled to
general relativity. Its action is given by Eqs. \Ref{action},
\Ref{xi}. Varying the action \Ref{action} gives the equation of
motion of the scalar field
    \begin{equation}\n{eqmo}
    \nabla^\alpha\nabla_\alpha\phi-V_\phi-\xi R\phi=0,
    \end{equation}
and the Einstein equations
    \begin{equation}\n{Einstein}
    G_{\mu\nu}=8\pi T_{\mu\nu},
    \end{equation}
where $V_\phi\equiv dV/d\phi$, $G_{\mu\nu}$ is the Einstein tensor,
and the stress-energy tensor is given by
      \bea\n{Tmn}
      T_{\mu\nu}&=&\nabla_{\mu}\phi\nabla_{\nu}\phi
      -\frac12g_{\mu\nu}(\nabla\phi)^2-g_{\mu\nu}V(\phi)
\nn
      &&+\xi\phi^2 G_{\mu\nu}-2\xi\left[\nabla_{\mu}(\phi\nabla_{\nu}\phi)
      -g_{\mu\nu}\nabla^{\lambda}(\phi\nabla_{\lambda}\phi)\right].
\nn
      \eea
By grouping all the dependence on $G_{\mu\nu}$ on the left-hand
side of the Einstein equations we can rewrite them in the
equivalent form
    \begin{equation}\n{Einsteineff}
    G_{\mu\nu}=8\pi T_{\mu\nu}^{\rm (eff)}
    \end{equation}
where \bea\n{Tmneff} T_{\mu\nu}^{\rm
(eff)}&=&\frac{1}{1-8\pi\xi\phi^2}
\ls\nabla_{\mu}\phi\nabla_{\nu}\phi
-\frac12g_{\mu\nu}(\nabla\phi)^2-g_{\mu\nu}V(\phi)
-2\xi\left(\nabla_{\mu}(\phi\nabla_{\nu}\phi)
-g_{\mu\nu}\nabla^{\lambda}(\phi\nabla_{\lambda}\phi)\right)\rs.
\nn \eea is an effective stress-energy tensor. $T_{\mu\nu}$ and
$T_{\mu\nu}^{\rm (eff)}$ are related as
\begin{equation}T_{\mu\nu}^{\rm (eff)}= \frac{T_{\mu\nu}-\xi\phi^2
G_{\mu\nu}}{1-8\pi\xi\phi^2}.
\end{equation}

For a static, spherically symmetric configuration, the metric can
be taken in the form \cite{Bron1,BS}
        \begin{equation}\n{metric}
        ds^2=-A(\rho) dt^2+\frac{d\rho^2}{A(\rho)}
        +r^2(\rho) (d\theta^2+\sin^2\theta\, d\varphi^2),
        \end{equation}
and $\phi=\phi(\rho)$. The nontrivial equations of the system
\Ref{Einsteineff} are
\begin{subequations}\n{nontriv}
\bea
2A\frac{r''}{r}+A'\frac{r'}{r}+A\frac{{r'}^2}{r^2}-\frac{1}{r^2}
&=&-8\pi\varepsilon^{\rm (eff)},\n{eq1}\\
A'\frac{r'}{r}+A\frac{{r'}^2}{r^2}-\frac{1}{r^2}&=&8\pi\sigma^{\rm
(eff)},
\n{fint}\\
\frac12 A''+A'\frac{r'}{r}+A\frac{r''}{r}&=&8\pi p^{\rm
(eff)},\n{eq3} \eea
\end{subequations}
where the prime denotes $d/d\rho$, and $\varepsilon^{\rm
(eff)}=-{T_0^0}^{\rm (eff)}$ is the effective energy density,
$\sigma^{\rm (eff)}={T_1^1}^{\rm (eff)}$ is the effective radial
pressure, and $p^{\rm (eff)}={T_2^2}^{\rm (eff)}={T_3^3}^{\rm
(eff)}$ is the effective transverse pressure. Using Eq.
\Ref{Tmneff} we find \bea
\varepsilon^{\rm(eff)}&=&\frac{1}{1-8\pi\xi\phi^2} \ls \frac12
A{\phi'}^2+V(\phi)-\xi A(\phi^2)''-\xi(\phi^2)'\(\frac12
A'+2A\frac{r'}{r}\)\rs,\\
\sigma^{\rm(eff)}&=&\frac{1}{1-8\pi\xi\phi^2} \ls \frac12
A{\phi'}^2-V(\phi)+\xi(\phi^2)'\(\frac12 A'+2A\frac{r'}{r}\)\rs,\\
p^{\rm(eff)}&=&\frac{1}{1-8\pi\xi\phi^2} \ls -\frac12
A{\phi'}^2-V(\phi)+\xi
A(\phi^2)''+\xi(\phi^2)'\(A'+A\frac{r'}{r}\)\rs. \eea Only two of
three equations of the system \Ref{nontriv} are independent: Eq.
\Ref{fint} is a first integral of the others. Certain combinations
of Eqs. \Ref{nontriv} give two independent equations in the
following form:\,\footnote{Eq. \Ref{set1} can be obtained as the
combination $A^{-1}[{\rm\Ref{eq1}}-{\rm\Ref{fint}}]$. To get Eq.
\Ref{set2} one should construct the following combination: $r^2[
{\rm\Ref{eq1}}+{\rm\Ref{fint}}-2{\rm\Ref{eq3}}]$, and then take
into account Eq. \Ref{set1}.} \bea
2\frac{r''}{r}&=&-\frac{8\pi}{1-8\pi\xi\phi^2}\ls
{\phi'}^2-\xi(\phi^2)''\rs,
\n{set1}\\
A(r^2)''-A''r^2-2&=&-\frac{8\pi\xi(\phi^2)'}
{1-8\pi\xi\phi^2}r^4\(\frac{A}{r^2}\)'. \n{set2} \eea Note that
the potential $V(\phi)$ does not enter into the above expressions.
The equation of motion of the scalar field, Eq. \Ref{eqmo}, in the
metric \Ref{metric} reads
\begin{equation}\n{seteqmo}
(r^2A\phi')'-r^2V_\phi-\xi r^2R\phi=0,
\end{equation}
with
\begin{equation}\n{R}
R=-A''-4A'\frac{r'}{r}-4A\frac{r''}{r}-2A\frac{{r'}^2}{r^2}+\frac{2}{r^2}.
\end{equation}
Given the potential $V(\phi)$ Eqs. \Ref{seteqmo}, \Ref{set1}, and
\Ref{set2} are a determined set of equations for the unknowns $r$,
$A$, and $\phi$. Note that in the case $\xi=0$ this system
reproduces the equations given in Ref. \cite{Bron1}.

It is worth to emphasize that the choice of a general spherically
symmetric metric in the form \Ref{metric} is convenient for a
number of reasons \cite{Bron1,BS}. In particular, as was shown in
Refs. \cite{Bron1,BS} for the minimal coupled scalar field, the
equation \Ref{set2} can be integrated. Fortunately, in the general
case with arbitrary $\xi$ one can integrate Eq. \Ref{set2} as
well. Really, using a new denotation
\begin{equation}\n{denote}
B=r^4\(\frac{A}{r^2}\)',
\end{equation}
we can rewrite Eq. \Ref{set2} as follows
\begin{equation}\n{eqB}
(fB)'=-2f.
\end{equation}
Integrating the last equation gives
\begin{equation}\n{inteqB}
B=-\frac{2}{f}\(\int^\rho f(\zeta)d\zeta-c_1\),
\end{equation}
where $c_1$ is an integration constant. Taking into account Eq.
\Ref{denote} and integrating once more we obtain
\begin{equation}A=r^2\(c_2+\int^\rho\frac{B(\zeta)}{r^4(\zeta)}d\zeta\),
\end{equation}
where $c_2$ is another integration constant.

\section{Wormholes and the kink-like configuration of a scalar field: Some
general
results}
Consider a symmetry-breaking potential $V(\phi)$ with two minima
at $\phi_1$ and $\phi_2$, respectively (see Fig.1a). We will
assume that the potential is positively definite, $V(\phi)\ge 0$,
and $V(\phi_1)=V(\phi_2)=0$, so that the values $\phi_1$ and
$\phi_2$ correspond to the real vacuum of the scalar field.

We will search such solutions of Eqs. \Ref{seteqmo}, \Ref{set1},
\Ref{set2} which describe a static spherically symmetric
traversable wormhole. This means that the metric functions
$r(\rho)$ and $A(\rho)$ should possess the following properties
(for details, see Ref. \cite{VisserBook}):
\begin{itemize}
\item $r(\rho)$ has a global minimum, say $r_{*}=r(\rho_{*})$, such that
$r_{*}>0$; the point $\rho_{*}$ is the wormhole's throat, and
$r_{*}$ is the throat's radius;
\item $A(\rho)$ is everywhere positive, i.e. no event horizons exist in the
spacetime;
\item the spacetime geometry is regular, i.e. no singularities exist;
\item $r(\rho)\to|\rho|$ and $A(\rho)\to {\rm const}$ at
$|\rho|\to\infty$; this guarantees an existence of two
asymptotically flat regions ${\cal R}_\pm:\ \rho\to\pm\infty$
connected by the throat.
\end{itemize}

As for the scalar field, we will suppose that it has a kink-like
configuration with nontrivial topological boundary conditions,
i.e., the scalar field is in one of the vacuum states, say
$\phi=\phi_1$, in the asymptotically flat region ${\cal R}_-$, and
it is in the other vacuum state, $\phi=\phi_2$, in ${\cal R}_+$;
in the intermediate region $-\infty<\rho<\infty$ the field
smoothly varies from $\phi_1$ to $\phi_2$ (see Fig.1b). Such the
field configuration represents a spherical domain wall (interface)
localized near the wormhole throat.

The theory described by the set of nonlinear second-order
differential equations \Ref{set1}, \Ref{set2} and \Ref{seteqmo} is
rather complicated and it is necessary to use numerical methods
for its complete study. Until numerical calculations we will
analyze some general properties of the theory. The crucial role in
our analysis will play the function $f(\phi)=1-8\pi\xi\phi^2$
whose behavior is determined by the values $\xi$, $\phi_1$ and
$\phi_2$. Let us consider various cases:

\vskip6pt\noindent{\bf 1.} $\xi=0$. In this case the action
\Ref{action} describes the theory of a scalar field {\em
minimally} coupled to general relativity. No-go theorems proven in
Refs. \cite{Bron1,BS} state, in particular, that wormhole
solutions are not admitted in this case. It will be useful to
reproduce here the proof of this statement: For $\xi=0$ Eq.
\Ref{set1} reduces to
\begin{equation}2\frac{r''}{r}=-8\pi{\phi'}^2.
\end{equation}
Since $r(\rho)\ge 0$ by its geometric meaning, the last equation
gives $r''\le 0$, which rules out an existence of regular minima
of $r(\rho)$, and hence an existence of wormhole solutions.

\vskip6pt\noindent{\bf 2.} $\xi<0$. In this case the function
$f^{-1}=(1-8\pi\xi\phi^2)^{-1}$ is regular (i.e., finite, smooth,
and positive) in the whole range of $\rho$, and so we can assert
that a wormhole solution is impossible. The proof of this
assertion is given in Refs. \cite{Bron-CC1,Bron-CC2}, where the
reader can find details. Here we shortly reproduce main points of
that proof. With this purpose we introduce a new metric $\bar
g_{\mu\nu}$ and a new scalar field $\psi$ using Wagoner's
\cite{Wagoner} conformal transformation
\bea
&&g_{\mu\nu}=F(\psi)\bar g_{\mu\nu},\quad F=\frac1{f},\nonumber\\
&&\frac{d\psi}{d\phi}=\pm\frac{\sqrt{|l(\phi)|}}{f(\phi)}, \quad
l(\phi)=f+\frac32\(\frac{df}{d\phi}\)^2. \n{trans} \eea Now one
can write the action \Ref{action} as follows:
\begin{equation}\n{newaction}
S=\int d^4x\sqrt{-\bar g}\ls \frac1{8\pi}\overline R -(\partial
\psi)^2- 2\overline V(\psi) \rs,
\end{equation}
where
\begin{equation}\overline V(\psi)=\frac{1}{f^2}V(\phi).
\end{equation}
One can see that the transformation \Ref{trans} removes the
nonminimal coupling expressed in the $\phi$-dependent coefficient
before $R$ in Eq. \Ref{action}, so that Eq. \Ref{newaction}
represents the action of the scalar field $\psi$ {\em minimally}
coupled to general relativity. Because the conformal factor
$F=f^{-1}$ is regular the theories \Ref{action} and
\Ref{newaction} are conformally equivalent, in particular, the
global structure of spacetimes $M[g]$ and $\overline M[\bar g]$ is
the same. No-go theorems proven in Refs. \cite{Bron1,BS} state
that the theory \Ref{newaction} does not admit wormhole solutions.
Hence we should conclude that wormholes are also impossible in the
theory \Ref{newaction} with $\xi<0$.

\vskip6pt\noindent{\bf 3.} $\xi>0$.

\vskip6pt\noindent{\bf 3a:} $|\phi_1|<(8\pi\xi)^{-1/2}$ and
$|\phi_2|<(8\pi\xi)^{-1/2}$ (i.e., $|\phi|$ is small);

\vskip6pt\noindent{\bf 3b:} $|\phi_1|>(8\pi\xi)^{-1/2}$,
$|\phi_2|>(8\pi\xi)^{-1/2}$ (i.e., $|\phi|$ is large), and
$\phi_1$, $\phi_2$ have the same sign (both $\phi_1$ and $\phi_2$
are positive or negative).

\vskip6pt\noindent Since, for the cases 3a and 3b, the conformal
factor $F=f^{-1}=(1-8\pi\xi\phi^2)^{-1}$ is regular in the whole
range of $\rho$, we can again assert that a wormhole solution is
impossible.

\vskip6pt\noindent{\bf 3c:} $|\phi_1|>(8\pi\xi)^{-1/2}$,
$|\phi_2|>(8\pi\xi)^{-1/2}$, and $\phi_1$, $\phi_2$ have opposite
signs (say $\phi_1<-(8\pi\xi)^{-1/2}$ and
$\phi_2>(8\pi\xi)^{-1/2}$).

\vskip6pt\noindent The case 3c represents the kink-like
configuration such that the scalar field smoothly interpolates
between two super-Planckian values $\phi_1<-(8\pi\xi)^{-1/2}$ and
$\phi_2>(8\pi\xi)^{-1/2}$. In this case the function
$f=1-8\pi\xi\phi^2$ becomes to be equal to zero at two points
$\rho_1$ and $\rho_2$, where
$|\phi(\rho_1)|=|\phi(\rho_2)|=\phi_0=(8\pi\xi)^{-1/2}$, and hence
the conformal factor $F=f^{-1}=(1-8\pi\xi\phi^2)^{-1}$ turns out
to be singular at these points. Referring to \cite{Bron-CC2}, we
say that the spheres $\rho=\rho_1$ and $\rho=\rho_2$ are
transition surfaces, $S_{\rm trans}$. Note that, though $F$ is
singular on $S_{\rm trans}$, the metric $g_{\mu\nu}$ can still be
regular if the metric $\bar g_{\mu\nu}$ specified by the conformal
transformation \Ref{trans} has an appropriate behavior on $S_{\rm
trans}$. Bronnikov \cite{Bron-CC1,Bron-CC2} called such the
situation a conformal continuation from $\overline M[\bar g]$ into
$M[g]$ and obtained some of properties of conformally continued
solutions. In particular, he pointed out that the no-go theorems
\cite{Bron1,BS} forbidding wormhole solutions in the theory
\Ref{newaction} cannot be directly transferred to the theory
\Ref{action} if the conformal factor $F$ vanishes or has a
singular behavior at some values of $\rho$. However, there are
still no general results allowing to assert without specifying $F$
either wormholes are admitted in the theory \Ref{action} or not.
To solve this problem in the case 3c we prove the following
theorem.

\vskip6pt{\em Theorem 1.} The field equations \Ref{set1},
\Ref{set2}, and \Ref{seteqmo} of the theory \Ref{action} do not
admit {\em regular}\, solutions where $\phi$ has the kink-like
configuration such that $\phi(\rho)$ is at least a $C^2$-smooth
function monotonically increasing from $\phi_1<-(8\pi\xi)^{-1/2}$
to $\phi_2>(8\pi\xi)^{-1/2}$ while $\rho$ runs from $-\infty$ to
$\infty$.

{\em Proof.} Due to the theorem's formulation the function
$f(\rho)=1-8\pi\xi\phi^2(\rho)$ turns into zero {\em only}\, at
two points, say $\rho_1$ and $\rho_2$, so that
$f(\rho_1)=f(\rho_2)=0$, and otherwise $f(\rho)\not=0$. Consider
Eq. \Ref{set2}. One can write it in the form \Ref{eqB} and then,
integrating, in the form \Ref{inteqB}. Rewrite Eq. \Ref{inteqB} as
follows:
\begin{equation}\n{theorem}
r^4\(\frac{A}{r^2}\)'=-2\frac{h}{h'},
\end{equation}
where $h(\rho)$ is a primitive function for $f(\rho)$, so that
$h'=f$. Note that the denominator in the right-hand side of Eq.
\Ref{theorem} is equal to zero at the points $\rho_1$ and
$\rho_2$, $h'(\rho_1)=h'(\rho_2)=0$. Suppose that the nominator
becomes also to be equal to zero at these points,
$h(\rho_1)=h(\rho_2)=0$, so that the ratio $h/h'$ remains to be
regular. Because $h$ is a smooth function being equal to zero at
the boundaries of the interval $(\rho_1,\rho_2)$, there exists a
point within the interval, $\rho_1<\rho_*<\rho_2$, where $h$ has
an extremum, so that $h'(\rho_*)=0$. Hence $f(\rho_*)=0$. We have
obtained a contradiction which proves the theorem.

\vskip6pt An immediate corollary of the theorem 1 is that the
theory \Ref{action} does not admit wormholes supported by the
field configuration 3c. Note also that the proof of the theorem
rests actually on a feature of the function $f(\rho)$ possessing
exactly two zeros. It is obvious that the theorem can be extended
to cases when $f(\rho)$ has an even number of zeros.

\vskip6pt\noindent{\bf 3d:} $|\phi_1|<(8\pi\xi)^{-1/2}$,
$|\phi_2|>(8\pi\xi)^{-1/2}$.

\vskip6pt\noindent The case 3d represents the kink-like
configuration such that the scalar field varies from a small value
$|\phi_1|<(8\pi\xi)^{-1/2}$ to a large value
$\phi_2>(8\pi\xi)^{-1/2}$. In this case there is just one and only
point, say $\rho_0$, where
$|\phi(\rho_0)|=\phi_0=(8\pi\xi)^{-1/2}$, and hence the conformal
factor $F=f^{-1}=(1-8\pi\xi\phi^2)^{-1}$ turns out to be singular
at this point. The theorem 1 proven for the configuration 3c does
not work now, and one may suppose that wormhole solutions
supported by the field configuration 3d could exist in the theory
\Ref{action}. Indeed, for the potential $V(\phi)$ being equal to
zero, such the solutions was found for the first time by Bronnikov
\cite{Bron73} (see also Ref. \cite{BV1}) for conformal coupling,
$\xi=1/6$, and more recently by Barcel\'o and Visser \cite{BV2}
for any $\xi>0$.

In the next section we will obtain a wormhole solution for the
symmetry-breaking potential $V(\phi)$.

\section{Wormhole solution}
\subsection{Model and analysis of boundary conditions}
Consider a `toy' symmetry-breaking potential
\begin{equation}\n{toyV}
V(\phi)=\frac{\lambda}{4}\ls(\phi-\bar\phi)^2-\frac{m^2}{\lambda}\rs^2,
\end{equation}
where $\lambda>0$, $m>0$, and $\bar\phi$ are some constants. The
minima of the potential \Ref{toyV} correspond to
\begin{equation}\phi_1=-\frac{m}{\sqrt{\lambda}}+\bar\phi, \quad
\phi_2=\frac{m}{\sqrt{\lambda}}+\bar\phi.
\end{equation}
Note that the configuration 3d is only possible provided
$\bar\phi\not=0$.

At present it will be convenient to introduce new dimensionless
variables and quantities
\bea
&&x=m\rho, \quad \tilde r(x)=mr(\rho), \quad
\eta(x)=\frac{\phi(\rho)}{\kappa}, \nonumber\\
&&\bar\eta=\frac{\bar\phi}{\kappa},\quad
\bar\eta_1=-1+\bar\eta,\quad \bar\eta_2=1+\bar\eta,\quad
\kappa=\frac{m}{\sqrt{\lambda}}. \eea Taking into account Eq.
\Ref{toyV} we rewrite the field equations \Ref{set1}, \Ref{set2}
and \Ref{seteqmo} in the dimensionless form: \bea
&&2\frac{r''}{r}=-\frac{8\pi\kappa^2}{1-8\pi\xi\kappa^2\eta^2}\ls
{\eta'}^2-\xi(\eta^2)''\rs, \\
&&A(r^2)''-A''r^2-2=-\frac{8\pi\xi\kappa^2(\eta^2)'}
{1-8\pi\xi\kappa^2\eta^2}\,r^4\(\frac{A}{r^2}\)',\\
&&(r^2A\eta')'-\xi
r^2R(x)\eta-r^2(\eta-\bar\eta)\ls(\eta-\bar\eta)^2-1\rs=0, \eea
with
$$
R(x)=-A''-4A'\frac{r'}{r}-4A\frac{r''}{r}-2A\frac{{r'}^2}{r^2}+\frac{2}{r^2},
$$
where the prime denotes $d/dx$. (Notice: For short hereafter we
drop a tilde over $\tilde r(x)$.) Resolve the last system in terms
of the second derivatives $r''$, $A''$, and $\eta''$: \bea
r''&=&\frac{r}{6\xi^2\eta^2+f}\ls-\frac12{\eta'}^2(1-2\xi)
-\xi\eta\eta'\(2\frac{r'}{r}+\frac{A'}{A}\)
+\xi\frac{1}{A}\eta\Delta\eta(\Delta\eta^2-1)\right. \nonumber\\
&&\left.-4\xi^2\eta^2\(\frac{{r'}^2}{r^2}+\frac{r'A'}{rA}-\frac1{r^2A}\)
+2\frac{\xi^3\eta^3\eta'}{f}\(2\frac{r'}{r}-\frac{A'}{A}\)\rs,
\n{der1}\\
A''&=&\frac{A}{6\xi^2\eta^2+f}\ls-{\eta'}^2(1-2\xi)
+2f\(\frac{{r'}^2}{r^2}-\frac{1}{r^2A}\)-8\xi\eta\eta'\frac{r'}{r}
+2\xi\frac1A\eta\Delta\eta(\Delta\eta^2-1)
\right.\nonumber\\
&&\left.+4\xi^2\eta^2\(\frac{{r'}^2}{r^2}-2\frac{r'A'}{rA}-\frac{1}{r^2A}\)
-8\frac{\xi^3\eta^3\eta'}{f}\(2\frac{r'}{r}-\frac{A'}{A}\)\rs,
\n{der2}\\
\eta''&=&\frac{1}{6\xi^2\eta^2+f}\ls-\eta'f\(2\frac{r'}{r}+\frac{A'}{A}\)
+\frac{f}{A}\Delta\eta(\Delta\eta^2-1)+3\xi\eta{\eta'}^2(1-2\xi)
\right. \nonumber\\
&&\left.-4\xi
f\eta\(\frac{{r'}^2}{r^2}+\frac{r'A'}{rA}-\frac1{r^2A}\)
+2\xi^2\eta^2\eta'\(2\frac{r'}{r}-\frac{A'}{A}\)\rs. \n{der3} \eea
where
\begin{equation}f=(8\pi\kappa)^{-2}(1-8\pi\xi\kappa^2\eta^2), \quad
\Delta\eta\equiv\eta-\bar\eta.
\end{equation}
Eqs. \Ref{der1}, \Ref{der2} and \Ref{der3} represent a system of
three ordinary second-order differential equations which has a
general solution depending, generally speaking, on six parameters.

Let us discuss the case of our interest. Suppose that there exists
a point $x=x_0$ where $\eta(x_0)=(8\pi\xi\kappa^2)^{-1/2}$, so
that $f(x_0)=0$; without loss of generality we can assume that
$x_0=0$. To ensure a regular behavior of the right-hand sides of
Eqs. \Ref{der1}, \Ref{der2} we should provide regularity of the
term $\frac1f\(2\frac{r'}{r}-\frac{A'}{A}\)$ at $x=0$. Hence we
obtain
\begin{equation}\n{rel1}
2\frac{r_0'}{r_0}-\frac{A_0'}{A_0}=0,
\end{equation}
where $r_0=r(0)$ and so forth. This formula determines a relation
between the values of the functions $r(x)$ and $A(x)$ and their
first derivatives at the point $x=0$. An additional relation can
be obtained by using Eq. \Ref{fint} which is the first-order
differential equation and plays a role of the constraint for
boundary conditions. At the point $x=0$ it gives
\begin{equation}\n{rel2}
\frac{r_0'}{r_0}=-\frac{\sqrt{8\pi}\,\kappa}{12\sqrt{\xi}\,\eta_0'A_0}
\(A_0{\eta_0'}^2-\frac12(\Delta\eta^2-1)^2\).
\end{equation}
Now the solution for $r(x)$, $A(x)$ and $\eta(x)$ being regular in
the vicinity of $x=0$ can be expanded in the
series:\,\footnote{Note that $r(x)$, $A(x)$ and $\eta(x)$ are not
only regular in the vicinity of $x=0$ but also analytic, i.e.,
they are infinitely differentiable at the point $x=0$. To prove
this one may differentiate Eqs. \Ref{der1}, \Ref{der2}, \Ref{der3}
together with the condition \Ref{rel1}.}
\bea
r(x)&=&r_0+r_0'x+\frac12r_0''x^2+\dots\, ,\nonumber\\
A(x)&=&A_0+A_0'x+\frac12A_0''x^2+\dots\, ,\nonumber\\
\eta(x)&=&\frac{1}{\kappa\sqrt{8\pi\xi}}+\eta_0'x+\frac12\eta_0''x^2
+\dots\, , \eea where only three coefficients, say $r_0$, $A_0$
and $\eta'_0$, are independent parameters; the others can be found
by using the relations \Ref{rel1} and \Ref{rel2} and Eqs.
\Ref{der1}, \Ref{der2} and \Ref{der3}.

It is necessary to stress that our analysis of boundary conditions
is local and based on the demand of regularity of a solution at
the critical point $x=0$ where the function $f(x)$ becomes to be
equal to zero. However this analysis does not answer the question
about an existence of a wormhole solution with the kink-like
configuration of the scalar field. In the next subsection we will
show that such the solutions do really exist.

\subsection{Numerical results}
Taking into account the above analysis of boundary conditions we
can solve the system of equations \Ref{der1}, \Ref{der2} and
\Ref{der3} numerically. To perform this in practice we must
specify the six parameters, where $\xi$, $\kappa$ and $\bar\eta$
are parameters of the model, and $r_0$, $A_0$ and $\eta'_0$ are
determining boundary conditions. Note that $\xi>0$ because the
configuration 3d is of our interest, $\kappa>0$ by definition, and
$r_0>0$ and $A_0>0$ by their geometric meaning. Though the local
boundary condition analysis did not reveal any restrictions for
$r_0$, $A_0$ and $\eta'_0$, we may however suppose that
topologically nontrivial solutions describing the kink-like
configuration of the scalar field could only exist for some
definite boundary conditions.%
\footnote{Analogously, the well-known kink solution
$y=\tanh{[(x-x_0)/\sqrt{2}]}$ of the nonlinear differential
equation $y''-y^3+y=0$ corresponds to the special choice of
boundary conditions with $y'_0=1/\sqrt{2}$.} In practical
calculations we will fix two of three parameters $r_0$, $A_0$,
$\eta'_0$, tuning then a value of third parameter in order to
obtain a solution with the kink-like field configuration. Some of
numerical results are shown in Figs. 2-4 for the following set of
parameters:
\bea\n{parameters}
&&\xi=\frac16,\quad \kappa=\frac1{\sqrt{8\pi}},\quad
\bar\eta=\frac{1}{\sqrt{8\pi\xi\kappa^2}},\nonumber\\
&& r_0=1,\quad A_0=36.321278,\quad  \eta'_0=0.2. \eea Let us
discuss the obtained results in details.

Graphs in Figs. 2-4 represent a numerical solution for the
functions $r(x)$, $A(x)$ and $\eta(x)$, respectively. The form of
graphs is nonsymmetric.\footnote{This dissymmetry of the solutions
just reflects the dissymmetry of the potential $V(\phi)$.} The
critical point $x=0$ separates two regions, $x<0$ and $x>0$, where
behavior of the solutions is qualitatively different. In the
region $x<0$ the functions $r(x)$, $A(x)$ and $\eta(x)$ are
monotonic and have the following asymptotical behavior at
$|x|\to\infty$:
\begin{equation}r(x)\approx k_1|x|,\quad A(x)\approx a_1-\frac{b_1}{|x|}, \quad
\eta(x)\approx \eta_1.
\end{equation}
where $k_1\approx 0.129$, $a_1\approx 70$, $b_1\approx 440$, and
$\eta_1=-1+(8\pi\xi\kappa^2)^{-1/2}\approx 1.4495$. While in the
region $x>0$ they have an oscillating component superposed on a
monotonic one, so that their asymptotics at $x\to\infty$ are
\begin{equation}r(x)\approx k_2x+p\,\sin\omega x,\quad A(x)\approx
a_2-\frac{b_2}{x}+q\,\sin\omega x, \quad
\eta(x)\approx\eta_2+s\,\sin\omega x
\end{equation}
where $k_2\approx 0.04$, $a_2\approx 770$, $b_2\approx 470000$,
and $\eta_2=1+(8\pi\xi\kappa^2)^{-1/2}\approx 3.4495$. Note that
the amplitudes $p$, $q$, $s$ and the frequency $\omega$ are
actually not constant, they depend on $x$ and are slowly
decreasing while $x$ increases; within the interval from 0 to 1000
their values can be approximated as $p\approx0.8$, $q\approx 10$,
$\omega\approx0.0769$.

It is necessary to emphasize that the solution obtained describes
a wormhole with the kink-like configuration of the scalar field.
Really, (i) the metric function $r(x)$ has the global minimum
$r_*\approx 0.9484$ at the point $x_*\approx 3.52$, which is the
wormhole's throat; (ii) $r(x)$ approaches to $k|x|$ and $A(x)$
approaches to a constant in the asymptotical regions
$|x|\to\infty$; this guarantees an existence of two asymptotically
flat regions of the spacetime; (iii) the scalar field $\eta$
varies from one vacuum state $\eta_1$ at $x=-\infty$ to the other
one $\eta_2$ at $x=\infty$; this represents the kink-like
configuration. The distribution of the scalar field energy density
$\varepsilon(x)$ is shown in Fig. 5. It is seen that it has a
narrow peak at $x\approx 6.6$ near the throat; on the left of the
peak the energy density is very fast decreasing; on the right the
energy density is also fast decreasing making small oscillations
around zero. Thus, the main part of the scalar field energy turns
out to be concentrated in the narrow spherical region near the
throat, and so we will call this region a {\em spherical domain
wall}.

As is well known, traversable wormholes as solutions to the
Einstein equations can only exist with exotic matter, for which
many of the energy conditions should be violated \cite{MT} (for
details, see also Refs. \cite{BV2,HV}). Let us discuss our
solution with this point of view. Consider the weak energy
condition (WEC) which reads
\begin{equation}G_{\mu\nu}v^{\mu}v^{\nu}=8\pi T^{\rm
(eff)}_{\mu\nu}v^{\mu}v^{\nu}\ge0,
\end{equation}
where $v^\mu$ is properly normalized timelike vector, $v^2=-1$.
For the static spherically symmetric configuration the WEC is
equivalent to the positivity of the effective energy density,
\begin{equation}\n{wec}
\varepsilon^{\rm (eff)}\ge 0.
\end{equation}
The graph for $\varepsilon^{\rm (eff)}$ is given in Fig. 6. It
shows, as was expected, that the condition \Ref{wec} is violated.

\section{Concluding remarks}
So, we have obtained that the theory \Ref{action}, \Ref{xi} with
the potential $V$ given by Eq. \Ref{toyV} admits solutions
describing a wormhole supported by the kink-like configuration of
the scalar field. Note that though the wormhole solution found
numerically in the previous section was only given for particular
values of parameters (see Eq. \Ref{parameters}), our calculations
have demonstrated that such solutions exist for a large scale of
parameter's values. Unfortunately, we have no reasonable
analytical estimations allowing to impose more exact restrictions
for a domain of admissible parameter's values. Some rough
restrictions for the parameters are dictated by the results of the
section III, where it was shown that a wormhole solution can only
exist for the configuration 3d. In particular, we have $\xi>0$.
Note that there are no restrictions for the parameter $r_0$ which
determines the throat's radius. In our consideration we have
specified $r_0=1$ and obtained the throat's radius
$r_*\approx0.9484$. In practice of numerical calculations we have
also used $r_0=10$ and $r_0=100$ with $r_*\approx10$ and
$r_*\approx100$, respectively.

The important problem which needs to be discussed is the stability
of solutions obtained. Recently, Bronnikov and Grinyok \cite{BG}
have shown that static spherically symmetric wormholes with the
nonminimally coupled scalar field with $V=0$ are unstable under
spherically symmetric perturbations. On the contrary, one may
expect that wormholes supported by the scalar field with the
symmetry-breaking potential $V$ would be stable because of the
topological stability of the kink-like field configuration. Of
course, this problem needs more serious consideration, and we
intend to discuss it in a following publication.

\section*{Acknowledgments}
We are grateful to Kirill Bronnikov for helpful discussions. S.S.
acknowledge kind hospitality of the Ewha Womans University. S.S.
was also supported in part by the Russian Foundation for Basic
Research grant No 99-02-17941. S.-W.K. was supported in part by
grant No. R01-2000-00015 from the Korea Science \& Engineering
Foundation.

\newpage
\begin{figure}
\includegraphics{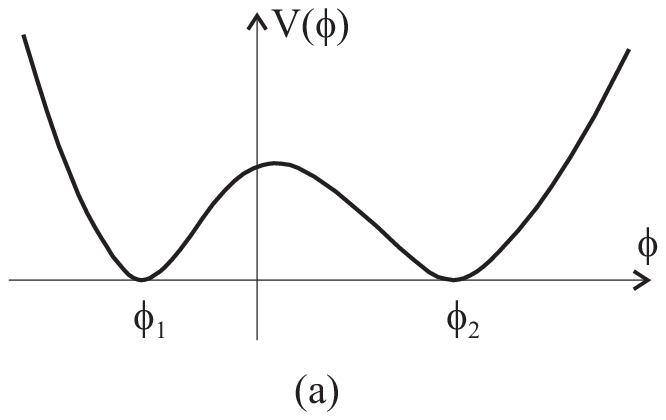}\hspace{1cm}
\includegraphics{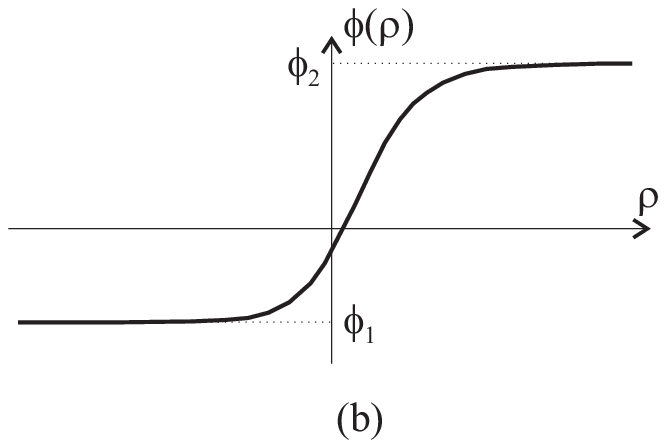}
\caption{(a) A general form of the symmetry-breaking
positive-definite potential $V(\phi)$. Two minima correspond to
two vacuum states of the scalar field; (b) The kink-like
configuration of the scalar field. The scalar field varies from
$\phi_1$ to $\phi_2$.} \label{f1}
\end{figure}

\begin{figure}\n{f2}
\includegraphics{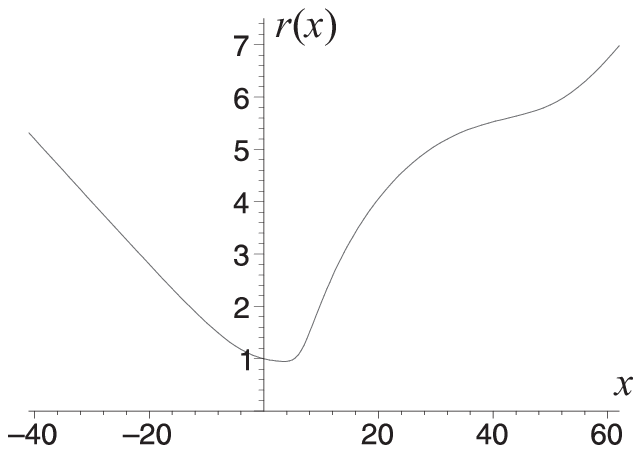}\hspace{1cm}
\includegraphics{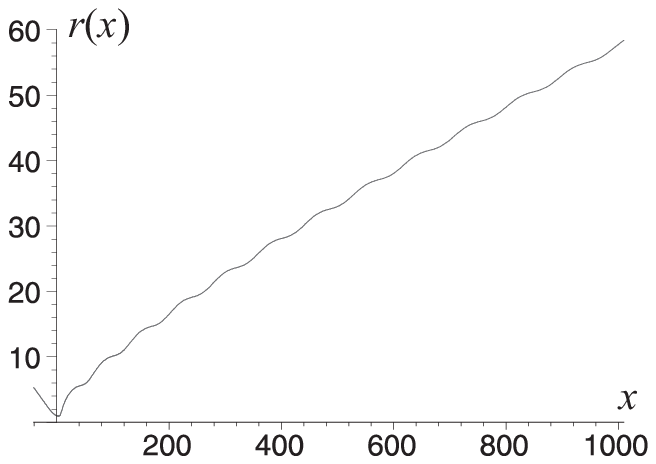}
\caption{The numerical solution for $r(x)$ in the small and large
scales; $\xi=1/6$, $\kappa=(8\pi)^{-1/2}$,
$\bar\eta=(8\pi\xi\kappa^2)^{-1/2}$.}
\end{figure}

\begin{figure}\n{f3}
\includegraphics{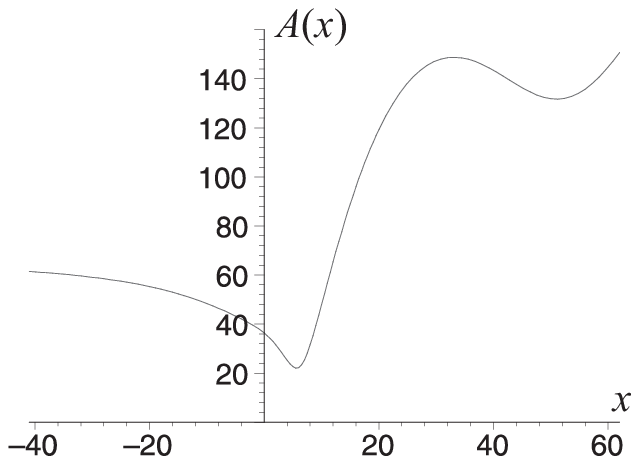}\hspace{1cm}
\includegraphics{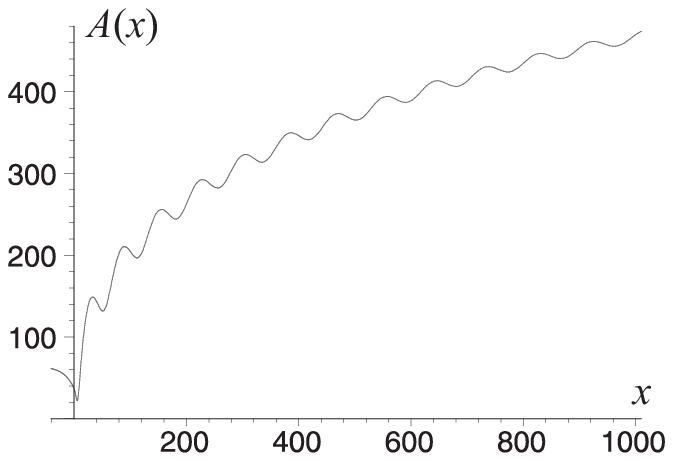}
\caption{The numerical solution for $A(x)$ in the small and large
scales; $\xi=1/6$, $\kappa=(8\pi)^{-1/2}$,
$\bar\eta=(8\pi\xi\kappa^2)^{-1/2}$.}
\end{figure}

\begin{figure}\n{f4}
\includegraphics{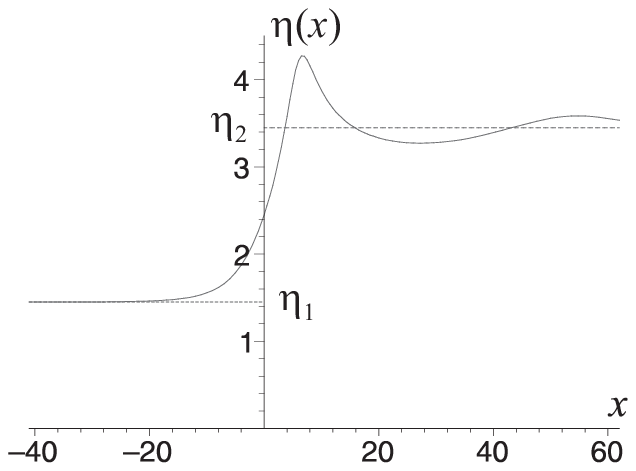}\hspace{1cm}
\includegraphics{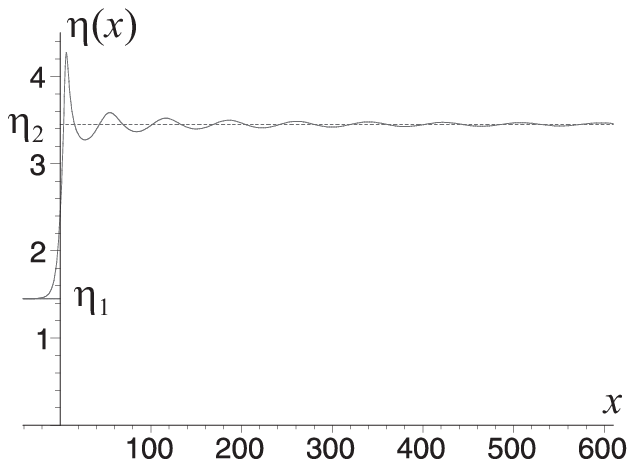}
\caption{The kink-like configuration of the scalar field. The
scalar field varies from
$\eta_1=-1+(8\pi\xi\kappa^2)^{-1/2}\approx 1.4495$ to
$\eta_2=1+(8\pi\xi\kappa^2)^{-1/2}\approx 3.4495$.}
\end{figure}

\begin{figure}\n{f5}
\includegraphics{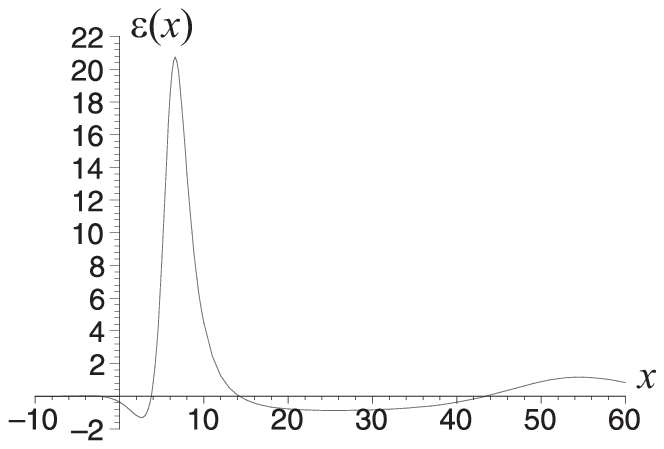}\hspace{1cm}
\includegraphics{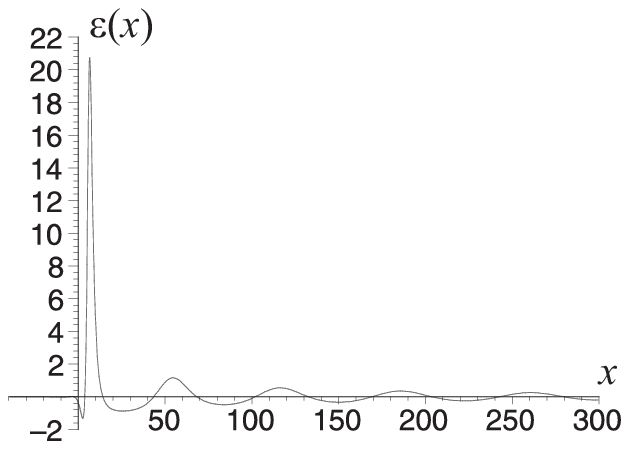}
\caption{The graph of the energy density $\varepsilon(x)=-T_0^0$
in the small and large scales. The highest peak at $x\approx 6.6$
corresponds to a spherical domain wall.}
\end{figure}

\begin{figure}\n{f6}
\includegraphics{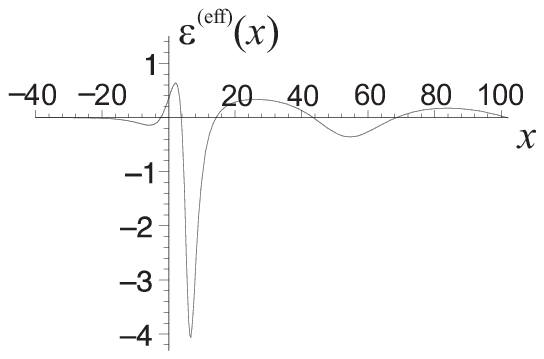}
\caption{The graph of the effective energy density
$\varepsilon^{\rm (eff)}(x)=-{T_0^0}^{\rm (eff)}$.}
\end{figure}


\begin{thebibliography}{99}
\bibitem{remark}
Among them are false vacuum black holes, cosmological and black
hole solutions to the Einstein-Yang-Mills equations, and many
others. However this field is very vast, and its review goes
beyond the present paper.
\end{thebibliography}
\end{document}